\documentclass[prl,twocolumn,showpacs,preprintnumbers,amsmath,amssymb]{revtex4}
\usepackage{epsfig}
\usepackage{amsmath}
\usepackage{graphicx}

\begin{document}

\title{Long-Range Connections in Transportation Networks}

\author{Matheus P. Viana}
\affiliation{Institute of Physics at S\~ao Carlos, University of
S\~ao Paulo, P.O. Box 369, S\~ao Carlos, S\~ao Paulo, 13560-970
Brazil}

\author{Luciano da F. Costa}
\email{costaldf@gmail.com}
\altaffiliation{National Institute of Science and Technology for Complex Systems, Brazil.}
\affiliation{Institute of Physics at S\~ao Carlos, University of
S\~ao Paulo, P.O. Box 369, S\~ao Carlos, S\~ao Paulo, 13560-970
Brazil}

\date{10th May 2010}

\begin{abstract}
Since its recent introduction, the small-world effect has been
identified in several important real-world systems. Frequently, it is
a consequence of the existence of a few long-range connections, which
dominate the original regular structure of the systems and implies
each node to become accessible from other nodes after a small number
of steps, typically of order $\ell
\propto \log N$. However, this effect has been observed in pure-topological
networks, where the nodes have no spatial coordinates. In this paper,
we present an alalogue of small-world effect observed in real-world
transportation networks, where the nodes are embeded in a
three-dimensional space. Using the multidimensional scaling method, we
demonstrate how the addition of a few long-range connections can
suubstantially reduce the travel time in transportation systems. Also,
we investigated the importance of long-range connections when the
systems are under an attack process. Our findings are illustrated for
two real-world systems, namely the London urban network (streets and
underground) and the US highways network enhanced by some of the main
US airlines routes.
\end{abstract}

\maketitle

\section{Introduction}

Long-range connections and their effects in systems modeled by complex
networks have been widelly studied in the last years. The most
important effect became knwon as \emph{small-world effect} as it
provides an elegant explanation for the Milgram's experiment of the
six degrees of separation \cite{Milgram1967}. The first suitable model
capable of explaining the small-world effect was reported by Duncan
J. Watts and Steven Strogatz \cite{Watts1998} in 1998, which has
motivated several applications to real-world problems. The small-world
model of Watts and Strogatz reveals how the inclusion of just a few
long-range connections into regular networks can drastically decrease
the network diameter (ie. the number of edges between two nodes) in
these networks. Considering that the displacements are done along the
shortest paths of the network, it is well-known
(e.g. \cite{Estrada2009}) that in a $d-$dimensional regular network,
where the nodes establish connections constrained by adjacency rules,
the average traveling time is of order $\tau \propto N^{1/d}/v$. Here,
$v$ correspond to the number of edges crossed per unit of time. By
adding a few number of long-range connections, it has been observed
that the average travel time descreases as $\tau \propto
\log(N)/v$. Although this approach is correct for many purely
topological systems, such as protein-protein
\cite{Barabasi2001}, \emph{WWW} \cite{Barabasi1999}, citations
\cite{Redner1998} and collaborations \cite{Watts1998} networks, we
should note that it cannnot be directly extended to several real-world
systems, where the Euclidean space and the displacement velocity play
a crucial rule in determining the transportation properties of the
network
\cite{Gastner2006,Hayashi2006,Gastner2007,Bebber2007,Hayashi2007}.

The above properties have already been explored by several preceding
works. For example, Hayashi and Matsukubo \cite{Hayashi2007} showed
how the addition of long-range connections can improve the robustness
of some embedded network models against intentional attacks. Recently,
G. Li et al. \cite{Li2010} also studied the effects of long-range
connections on regular lattices. The authors considered the addition
of long-range links between nodes $i$ and $j$ with probability
$p_{i,j}\propto r_{i,j}^{-\gamma}$, where $r_{i,j}$ is the Manhattan
distance between the nodes and $\gamma>0$. Their results indicate that
the optimal transport occurs when $\gamma=d+1$ for a $d$-dimensional
system, independently of the navigation strategy adopted.

\begin{figure}[!b]
\begin{center}
    \includegraphics[width=0.8\linewidth]{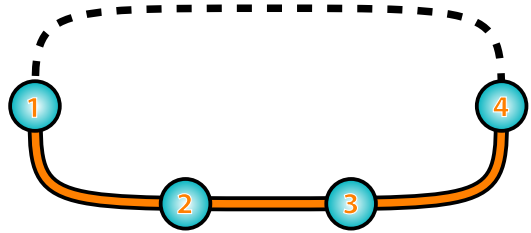}
    \caption{Sample network with three short-range connections and one
    long-range connection (dashed line). If we want to reach the node
    4 departing from 1, we can choose two shortest paths:
    $\{1,2,3,4\}$, which uses only short-range connections and
    $\{1,4\}$, which uses the long-range connection. In the point of
    view of the time spended to travel, there is no difference between
    the two paths if displacement velocities are the same along the
    two types of edges. In the other hand, if we can move faster in
    the long-range connection, we should choose the second path.}
    \label{fig:sample}
\end{center}
\end{figure}

\begin{figure*}[!htb]
\begin{center}
    \includegraphics[scale=0.6]{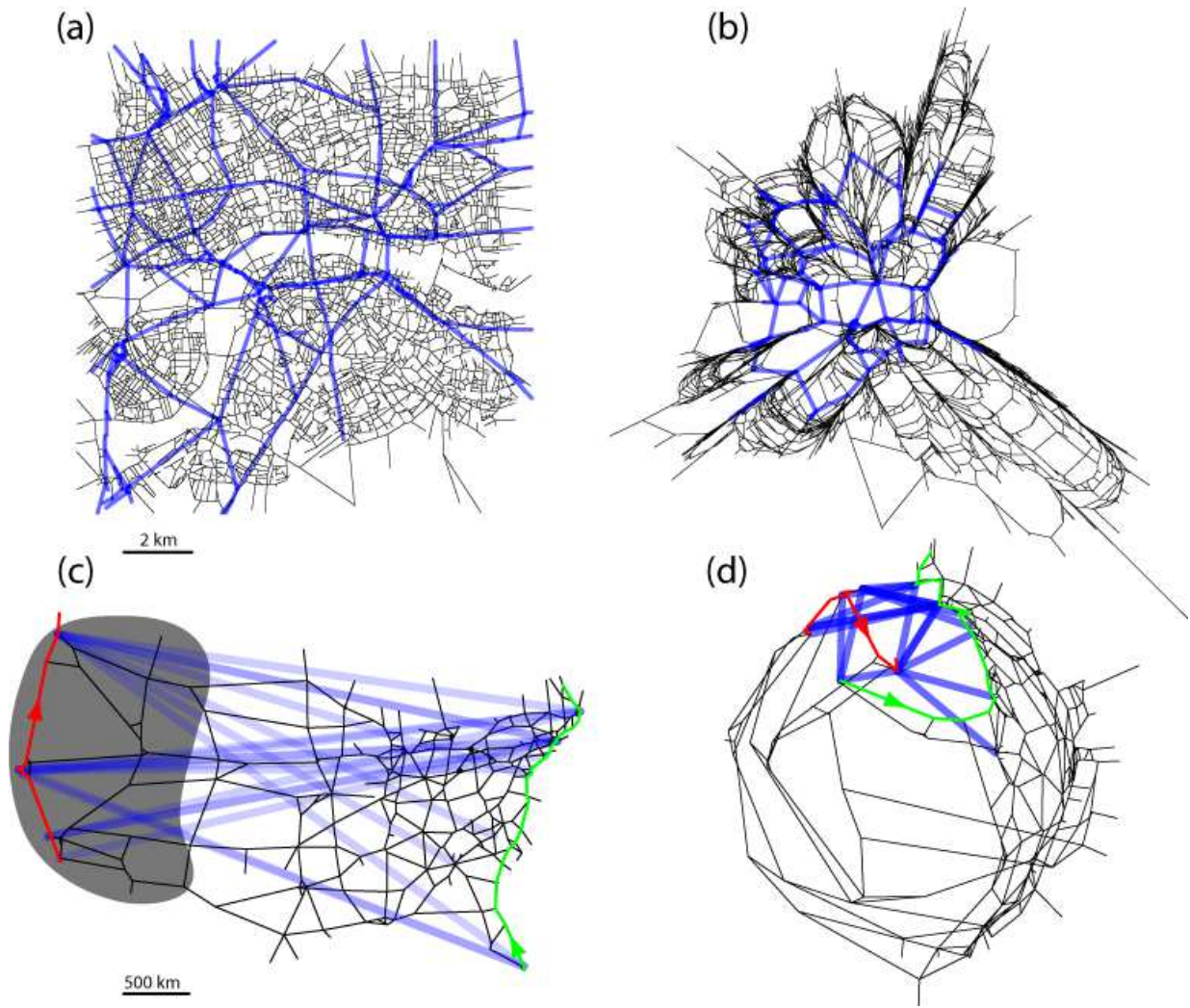}
    \caption{(a) The network of London streets and respective underground
    system (the latter in blue). (b) Layout obtained by multidimensional
    scaling. (c) Network of US Interstate Highway system enhanced by
    twenty of the most important links US Airlines links (in blue). (d)
    Layout obtained by multidimensional scaling. The adoped values of
    $\alpha$ were $\alpha=3$ for LSN and $\alpha=6$ for USHN. The
    red and green lines were placed in (c) and (d) just to identify
    the final positions on the west and east coast. Observe that the
    length of an edge in (b) and (d) is proportional to the time
    required to cross that edge.}
    \label{fig:mds}
\end{center}
\end{figure*}

In order to illustrate the importance of the space on the transport
properties, we show in Figure \ref{fig:sample} a simple embeded network with
three short-range connection and one long-range connection. If we want
to reach node 4 after departing from 1, we can choose two different
shortest paths (ie. paths with the minimum total length):
$\{1,2,3,4\}$ or $\{1,4\}$. Observe that the first path only uses
short-range connections, while the second option uses a long-range
link. It is clear that if the displacement velocity is fixed for both
types of connections, there is no advantage to use the first or second
option. However, as we will see, a substantially different result can
be obtained when we consider different displacement velocities for
long and short-range connections.

Commonly, real transportation networks are generalizations of the
simple situations discussed above. Real networks are embeded in
three-dimensional spaces and display regular properties, in which each
node is connected to a few number of neighboors through short-range
connections. This strategy reduces the building cost of real-world
structures, which have their costs proportional to the total length of
the system \cite{Flammini2008}. In addition, the network is enhanced by
a few number of long-range links that spans the space. The key point
here is to consider that the displacement velocity through the
short-range and long-range connections are different. While the
displacement along the short-range connections occurs at velocity
$v_s$, in long range-connection this velocity is $v_l = \alpha v_s$,
with $\alpha > 1$.

In the current paper we study two important real-world systems
characterized by the features discribed above, ie. they have two
types of connections with respectively different displacement
velocities. The systems that will be considered are (a) the network of
streets of London plus the respective underground system and (b) the
US highway plus some of the main US airlines connections. We will use
the multidimensional scaling in order to visualize the effect of the
long-range and will then quantify the importance of these connections
on the transportation properties as well. 

This paper is organized as follow: we start by presenting and
discussing the networks used here. Next, the multidimensional scaling
is used to visualize the effects of the network topology.  We then use
an attack dynamics in order to quantify the importance of the
long-range connections as compared to the short-range ones. Finally,
we present the main conclusions and prospects for further
investigation.

\section{Description of the data}

In this section we will how both networks used in this paper were
constructed: (a) the network of streets of London plus the respective
underground system and (b) the US highway plus some of the main US
airlines connections.

\subsection{London Streets Network (LSN)}

The central region of London, corresponding to about 13km$\times$13km,
was mapped into a network where each node corresponds to the
confluence of two or more streets. Also, the underground system of
London respective to the same region was then appended into this
network. Each underground station was replaced by its closer node from
the respective street network. For this network (henceforth called
LSN), we have $\alpha
\approx 3$ and $3\%$ of the total number of edges correspond to
long-range connections. The final network contains 5963 nodes and it
has an average degree of 2.81. Figure
\ref{fig:mds}(a) shows the final version of this network, where
the long-range-connections are depicted in blue. Observe that, for
this network, the long-range connections are 5 times larger than
short-range connections, in the average.

\subsection{US Highways Network (USHN)}

The sencond network (USHN) considered in this paper was built using
the American highway system enhanced by twenty of the most important
airlines connections. The importance of the airline connections was
quantified according the number of passengers that they transport. In
this network, the confluences of two or more highways were mapped into
nodes. Two nodes are linked if a highway connects them. Moreover, the
extremities of the airline connections were replaced by the closest
nodes from the highway system. For this network, we have
$\alpha\approx 6$ and the fraction of edges corresponding to
long-range-connections is $3\%$, again. The final version of this
network is showed in Figure \ref{fig:mds}(b) and it contains 428 nodes
with an average degree of 3.15. It is interesting to observe that
almost all long-range connection tend to link the the west coast to
the east coast. In average, we observed in this network, long-range
connections 40 times larger than short-range connections.

\begin{figure}[!t]
\begin{center}
    \includegraphics[width=\columnwidth]{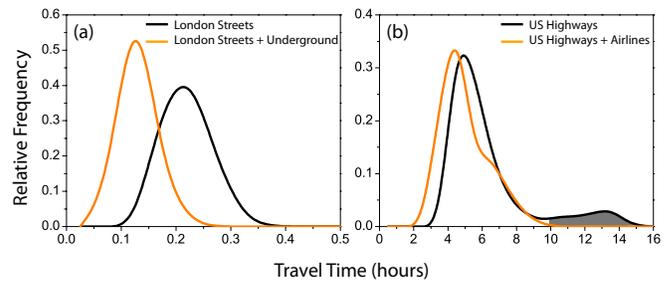}
    \caption{Effect of the long-range connection in the distribution
    of the traveling time for the (a) LSN and (b) USHN. In both cases,
    the distributions were shifted to left when the long-range
    connection were considered (orange lines). As we expected, this
    result indicates that the time required to travel between two
    nodes decreases in the presence of the long-range connections. In
    (a) the average value decreases 42\% while in (b) it decreases
    20\%. Observe that these results are affected by the number of
    long-range connections considered in each system. The
    average travel time reduction per long-range connection is about
    0.14\% in (a) and 2\% in (b).}  \label{fig:dmtt}
\end{center}
\end{figure}

\section{Visualizing the effect of the long-range connections}

We applied the classical mutidimensional scaling
~\cite{Kruskal1964,Berthold2003} on the networks in order to visualize
the effect of the long-range connections. This technique provides a
powerfull way for obtaining the nodes positions from a dissimilarity
matrix. We denote by $\tau_{i,j}$ the dissimilarity between the nodes
$i$ and $j$. In our case, this dissimilarity corresponds to the
minimum traveling time to reach the destination node $j$ after
departing from $i$. In order to evaluate the value of $\tau_{i,j}$
each edge $(i,j)$ of the network received an weight corresponding to
$\tau_{i,j} = \ell_{i,j}/v_{i,j}$, where $\ell_{i,j}$ is the Euclidean
distance between $i$ and $j$ and $v_{i,j}=v_s$ if $(i,j)$ is a
short-range connection or $v_{i,j} = \alpha v_s$ if $(i,j)$ is a
long-range connection. The dissimilarity matrix is defined as the
$N\times N$ symmetric matrix $\mathbf{T}$, which has elements
$\tau_{i,j}$. Now, the following matrix is obtained from $\mathbf{T}$:

\begin{equation}\label{eq1}
  \mathbf{B} = -\frac{1}{2} \left( \mathbf{I} - \frac{1}{N} \mathbf{U} \mathbf{U}^T \right) \mathbf{T}' \left( \mathbf{I} - \frac{1}{N}\mathbf{U} \mathbf{U}^T \right),
\end{equation}

where $\mathbf{U}$ is a vector $N \times 1$ whose elements are all
equal to one, $\mathbf{I}$ is the $N\times N$ identity matrix and
$\mathbf{T}'$ is a matrix whose elements are equal to the square of
the elements of $\mathbf{T}$, ie. $\tau '_{i,j} = \tau^2_{i,j}$. The
eigenvalues of $\mathbf{B}$ are then identified, and only those which
are larger than zero are considered in order to build the next matrix,
$\mathbf{E}$. Then, these $n \leq N$ eigenvalues are sorted in
decreasing order, yielding the sequence $\lambda_1 > \lambda_2 >
\ldots > \lambda_n > 0$.  The respective eigenvectors are stacked as
columns of a matrix $\mathbf{E}$ with dimension $N \times n$.  The
coordinates of the $N$ nodes can now be obtained, up to a rigid body
transformation, as:

\begin{equation}
\mathbf{X} = \mathbf{E} \left(
        \begin{array}{cccc}
          \sqrt{\lambda_1} & 0 & \ldots & 0 \\
          0 & \sqrt{\lambda_2} & \ldots & 0 \\
          \hdots & \hdots & \hdots & \hdots \\
          0 & 0 & \ldots & \sqrt{\lambda_n} \\
        \end{array}
      \right).
\end{equation}

The dimension of the final coordinates is approximately given by the
number of non-null eigenvalues, $n$. The higher the number of
constrainments of the dissimilarity matrix, the larger is the number
of dimensions required. Here we considered only two first eigenvectors
to visualize the final aspect of the transportation networks. The
results are showed in Figure \ref{fig:mds}(b) and Figure
\ref{fig:mds}(d) for LSN and USHN, respectvelly.

It is clear from the Figure \ref{fig:mds}(b) that the peripheral
regions of London were brought close to the central region. For the
USHN case - Figure \ref{fig:mds}(d), one can observe a folding effect
bringing together the west and the east coast over the US map. In both
these figures, the edge lengths are proportional to the traveling time
required to cross that respective edge. Thus, we expected that the
travel time averaged over all nodes of each network (\emph{average
travel time}) has decreased as a consequence of the addition of
long-range connections. This was confirmed by the results shown in
Figure \ref{fig:dmtt}, where we consider the distribution of the
travel time for both networks with and without the long-range
connections. It is possible to observe that in both cases the
traveling time was significantly decreased. Moreover, in the case of
the USHN, we can also note that the inclusion of the long-range
connections had a strong effect on the secondary peak of the time
travel distribution (gray region of Figure
\ref{fig:dmtt}(b)), meaning that the access to several nodes was
improved. We verifyed that these nodes, which had they access
improved, belong to the west coast and they are identified by the gray
region in Figure \ref{fig:mds}(c).

\section{Quantifying the importance of the long-range connections}

\begin{figure}[!htb]
\begin{center}
    \includegraphics[width=\columnwidth]{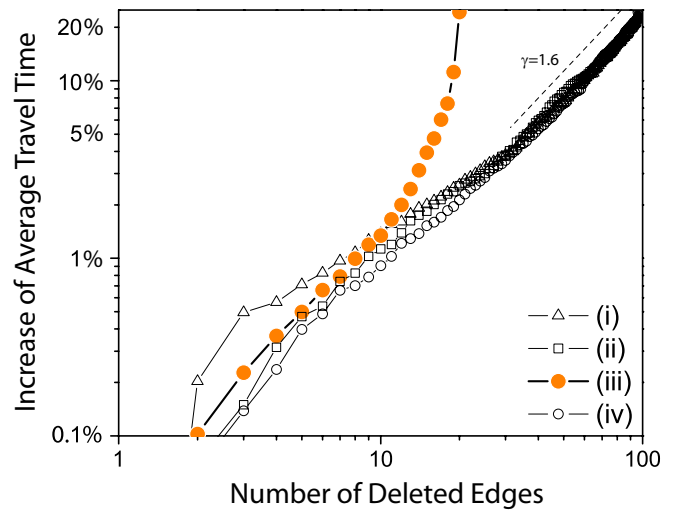}
    \caption{Behavior of the average travel time during the edges
    deletion process. The edges were deleted using four different
    strategies. In the first case, we considered the (i) US highways
    network without long-range connections and deleted its edges one
    by one (triangles). Next we considered the US highways network
    with long-range connections and deleted the (ii) short-range
    connections (squares), (iii) long-range connections (filled
    circles) and (iv) random connections (open circles). After each
    deletion, the new average travel time was re-evaluated. Also, along the
    deletion process, we chose edges that kept the network
    connected. The y-axis correspond to the percentual increase of the
    average travel time.}  \label{fig:attack}
\end{center}
\end{figure}

In order to investigate how the long-range connections can contribute
to improve the flow in transport networks, we performed a deletion
process under the USHN. At each time step, we chose a random edge of
the considered network and then checked if its respectively deletion
kept the network connected. If that was true, we deleted this edge and
re-evaluated the average traveling time. Next, a new edge was chosen,
and so on. This process finishes when no edges can be deleted. The
following configurations were considered: (i) deletion of the
short-range connections from the highways network without long-range
connections; (ii) deletion of the short-range connections from the
highways network with long-range connections; (iii) deletion of the
long-range connections from the highways network with long-range
connections; and (iv) deletion of the short-range and long-range
connections from the highways network with long-range connections. The
results are shown in Figure \ref{fig:dmtt}. As one can see, the
configurations (i), (ii) and (iv) have similar results and they
converge to a linear function with slope $\gamma=1.6$ when more than
30 edges are deleted. This behavior was not observed in the case
(iii), which diverged quickly while the percentual increase of the
average travel time reached about $25\%$ when twenty long-range
connections were deleted. For the other cases, it increased only
$2.5\%$ in the average. In addition, we can observe in Figure
\ref{fig:attack} that about a hundred deleted edges are required in
order to cause the same damage in (i), (ii) and (iv) as in
(iii). These results show that the long-range connection are able to
substantially improve the resilience in networks, but at the same
time, they can be considered as potential targets for preferential
attacks.

\section{Conclusions}

All in all, the results showed here lead us to believe that the
reduction effect of the travel time observed above can be considered
as an analogue of the small-world effect for the transportation
networks. For these networks, the nodes have a very-well defined
positions on space and tend to be linked with the neighboors trought
short-range connections in order to minimize the building cost. If a
few number of long-range connections are introduced in the network,
overcomming the original regular structure, the time spended to travel
in this network decreases significantly. It is important to note that
this result becomes valid when the displacement velocities along the
short-range ($v_s$) and long-range ($v_l$) connections are
different. 

We studied two real transport networks: (i) the London streets
networks plus the respective underground system (LNS) and (ii) the US
highway system enhanced by some of the main airline connections
(USHN). For these systems, we considered $v_l=\alpha v_s$, with
$\alpha=3$ for LSN and $\alpha=6$ for USHN.  By using the
multidimensional scaling methodology, we showed how the long-range
connections change the effective geography of the networks, bringing
together regions that are far away in the maps. These visual results
provided by the multidimensional scaling were confirmed by the
left-shift observed in the time traveling distributions, meaning that
the the addition of the long range-connections descreased the time
spent while traveling in the networks. 

The importance of the long-range connections againts the short-range
ones was quantifyed b yusing an edge deletion process, in which, at
each time step, an edge of the networks was deleted and the time
travel was re-evaluated. We observed that when the deletion is
performed only over the long-range connections, the time travel
diverges quickly, while it has a linear behavior when the other
deletion strategies are considered.

\begin{acknowledgments}
Luciano da F. Costa is grateful to FAPESP (05/00587-5) and CNPq
(301303/06-1 and 573583/2008-0) for the financial support. M. P.
Viana is grateful to FAPESP sponsorship (proc. 07/50882-9).
\end{acknowledgments}

\bibliography{viana10lrc}

\end{document}